\documentclass{Pos}

\usepackage{dsfont}
\usepackage{slashed}
\usepackage{booktabs}
\usepackage{graphicx}
\usepackage{float}

\newcommand{\Op}{\mathcal{O}} 
\newcommand{\eins}{\mathds{1}} 
\newcommand{\be}{\begin{equation}}
\newcommand{\ee}{\end{equation}}
\newcommand{\bea}{\begin{eqnarray}}
\newcommand{\eea}{\end{eqnarray}}
\newcommand{\beq}{\begin{eqnarray}}
\newcommand{\eeq}{\end{eqnarray}}

\title{$\Delta$ electromagnetic form factors and quark transverse charge densities from lattice QCD}

\ShortTitle{$\Delta$ e.m. form factors and transverse charge densities}

\author{\speaker{C. Alexandrou}\\
       Department of Physics, University of Cyprus, P.O. Box 20537,
 1678 Nicosia , Cyprus, and\\ Computation-based Science and Technology Research Center, Cyprus Institute, 15, Kypranoros Str., 1645 Nicosia, Cyprus \\
        E-mail: \email{alexand@ucy.ac.cy}}
\author{T. Korzec\\
        Department of Physics, University of Cyprus, P.O. Box 20537, 1678 Nicosia, Cyprus\\
        E-mail: \email{korzec@ucy.ac.cy}}
\author{G. Koutsou\\
Department of Physics, University of Cyprus, P.O. Box 20537, 1678 Nicosia, Cyprus,\\
Bergische Universit\"at Wuppertal, Fachbereich Physik, 42097 Wuppertal, Germany,\\
JSC and IAS, FZ J\"ulich, 52425 J\"ulich, Germany\\
   E-mail: \email{i.koutsou@fz-juelich.de}}
\author{C. Lorc\'e, V. Pascalutsa and M. Vanderhaeghen\\
       Institut f\"ur Kernphysik, Johannes Gutenberg-Universitaet Mainz,   
 J.J. Becher-Weg 45, D-55099 Mainz
 Germany\\
        E-mail: \email{marcvdh@kph.uni-mainz.de}}
\author{J. W. Negele\\
Center for Theoretical Physics, Laboratory for Nuclear Science
  and Department of Physics, Massachusetts Institute of Technology, Cambridge,
Massachusetts 02139, U.S.A.}
\author{A. Tsapalis\\
Institute of Accelerating Systems and Applications, University of
Athens, Athens, Greece and\\
Hellenic Naval Academy, Hatzikyriakou Ave., Pireaus, GR 18539, Greece\\
Email:\email{a.tsapalis@iasa.gr}}
\abstract{We discuss the techniques  to extract  the
          electromagnetic $\Delta$ form factors in Lattice QCD.
We evaluate these form factors using  dynamical fermions
          with  smallest pion mass of about 
         350 MeV.
  We pay particular attention to the  extraction of the electric 
quadrupole form factor that  
          signals a deformation of the $\Delta$.         
 The magnetic moment of the $\Delta$ 
is extrapolated using a chiral effective field theory.
 Using the form factors we evaluate   
          the transverse density distributions  in the infinite
          momentum frame showing deformation in  the $\Delta$.
 }

\FullConference{6th International Workshop on Chiral Dynamics, CD09\\
		July 6-10, 2009\\
		Bern, Switzerland}

\begin{document}

\section{Introduction}

Electromagnetic form factors probe 
the structure of hadrons, yielding information on their size, shape and magnetization.
While the nucleon form factors and the $N\to\Delta$ transition
form factors have been studied quite thoroughly
both experimentally and on the lattice~\cite{Alexandrou:2006ru}, much less
has been done for the $\Delta$ form factors. Experiments are 
notoriously difficult due to the short mean life time of the
$\Delta$ of only about $6\times 10^{-24} s$. Nevertheless the magnetic 
moments of the $\Delta^+$~\cite{Kotulla:2002cg} and 
$\Delta^{++}$~\cite{LopezCastro:2000cv,Yao:2006px} have been 
measured. 
Lattice Quantum Chromodynamics (QCD) provides a well-defined framework to directly calculate the $\Delta$ form factors from
the fundamental theory of  strong interactions. 
A primary motivation for this work is to understand the role of deformation 
in hadron structure. For hadrons a classical  non-relativistic description
is not adequate and the definition of shape has to be refined~\cite{RMP}.
A suitable framework to discuss the shape of such systems is to define 
transverse charge distributions in the infinite momentum frame. 
Thus, a major achievement of this work is the development 
of lattice methods with sufficient precision to extract  the 
electric quadrupole form factor and connect it in a rigorous manner
to a well-defined charge density distribution
so that the shape of the $\Delta$ can be discussed.
In order to evaluate the $\Delta$ electromagnetic 
(e.m.) form factors to the required accuracy,  we  isolate
the sub-dominant electric 
quadrupole form factor (FF) from the dominant electric and magnetic FFs.
 This is crucial  since without the construction of an optimized source for the
sequential propagator the electric quadrupole form factor  can not be extracted to the desired precision.
We note that this 
increases the computational cost since additional inversions are needed.
These techniques were first tested in quenched QCD~\cite{Alexandrou:2007dt}. Here
we present results in the quenched approximation, using two degenerate flavors ($N_F=2$)
 of Wilson fermions and in a mixed action approach where domain wall valence quarks on
a staggered $N_F=2+1$ sea are used.

\section{The transverse charge densities for a spin-3/2 particle}
\label{sec1}

We define a quark charge density for a spin-3/2 particle, such as the 
$\Delta(1232)$, in a state of definite light-cone helicity $\lambda$, 
by the Fourier transform~\cite{Alexandrou:2008bn,Alexandrou:2009hs},
\be
\rho^\Delta_{\lambda} (b) 
\equiv \int \frac{d^2 \vec q_\perp}{(2 \pi)^2} 
\, e^{- i \, \vec q_\perp \cdot \vec b} \, \frac{1}{2 P^+} 
\langle P^+, \frac{\vec q_\perp}{2}, \lambda 
| J^+ | P^+, \frac{- \vec q_\perp}{2}, \lambda  \rangle  
= \int_0^\infty \frac{d Q}{2 \pi} Q \, 
J_0(Q b) \, A_{\lambda \lambda}(Q^2),
\label{eq:dens2}
\ee
in the Breit frame, where $A_{\lambda\lambda}(Q^2)$ are helicity amplitudes and 
 $Q^2=-q^2=(p_f-p_i)^2$ with $p_i$ and $p_f$ the initial and final
momentum. 
The two independent quark charge densities for a spin-3/2 state 
of definite helicity are given by 
$\rho^\Delta_{\frac{3}{2}}(b)$ and 
$\rho^\Delta_{\frac{1}{2}}(b)$. 
Note that for a point-like particle, the `natural' values 
 lead to $A_{\frac{3}{2} \frac{3}{2}} (Q^2) = 1$~\cite{Alexandrou:2009hs}, 
implying 
$
\rho_{\frac{3}{2}}(\vec b) = \delta^2(\vec b).  
$
The above charge densities provide us with two combinations of
the four independent $\Delta$ FFs. To get information from 
the other FFs, we 
consider the charge densities in a spin-3/2 state with transverse spin.  
We denote this transverse polarization direction by 
$\vec S_\perp = \cos \phi_S \hat e_x + \sin \phi_S \hat e_y$, and 
the $\Delta$ spin projection along the direction of $\vec S_\perp$ by 
$s_\perp$. 
We can  define the charge densities in a spin-3/2 state 
with transverse spin $s_\perp$ as~: 
\begin{eqnarray}
\rho^\Delta_{T \, s_\perp}(\vec b) 
&\equiv& \int \frac{d^2 \vec q_\perp}{(2 \pi)^2} \,
e^{- i \, \vec q_\perp \cdot \vec b} \, \frac{1}{2 P^+} 
\langle P^+, \frac{\vec q_\perp}{2}, s_\perp  
\,|\, J^+(0) \,|\, P^+, -\frac{\vec q_\perp}{2}, s_\perp   \rangle. 
\label{eq:dens3}
\end{eqnarray}
By expressing the transverse spin basis in terms of the helicity
basis for spin-3/2~\cite{Alexandrou:2009hs}
and  working out the Fourier transform in Eq.~(\ref{eq:dens3}) 
for the two cases where $s_\perp = \frac{3}{2}$ 
and $s_\perp = \frac{1}{2}$
one obtains:
\newpage
\beq
\rho^{\Delta}_{T\, \frac{3}{2}}(\vec b)
&=&\int_0^{+\infty}\frac{d Q}{2\pi}\,Q\,
\Big[ J_0(Qb) \, \frac{1}{4} \left(A_{\frac{3}{2} \frac{3}{2}} 
+ 3 A_{\frac{1}{2} \frac{1}{2}} \right) 
- \sin(\phi_b-\phi_S) \, J_1(Qb) \,
\frac{1}{4} \left(2 \sqrt{3} A_{\frac{3}{2} \frac{1}{2}} 
+ 3 A_{\frac{1}{2} -\frac{1}{2}} \right) 
\nonumber \\
&-&\cos[2(\phi_b-\phi_S)] \, J_2(Qb) \,
\frac{\sqrt{3}}{2} A_{\frac{3}{2} -\frac{1}{2}}
+ \sin[3(\phi_b-\phi_S)] \, J_3(Qb) \,
\frac{1}{4} A_{\frac{3}{2} -\frac{3}{2}}
\Big], 
\label{eq:dens4} 
\end{eqnarray}
and
\begin{eqnarray}
\rho^{\Delta}_{T\, \frac{1}{2}}(\vec b)
&=&\int_0^{+\infty}\frac{d Q}{2\pi}\,Q\,
\Big[ J_0(Qb) \, \frac{1}{4} \left(3 A_{\frac{3}{2} \frac{3}{2}} 
+ A_{\frac{1}{2} \frac{1}{2}} \right) 
- \sin(\phi_b-\phi_S) \, J_1(Qb) \, 
\frac{1}{4} \left(2 \sqrt{3} A_{\frac{3}{2} \frac{1}{2}} 
- A_{\frac{1}{2} -\frac{1}{2}} \right) 
\nonumber \\
&+&\cos[2(\phi_b-\phi_S)] \, J_2(Qb) \, 
\frac{\sqrt{3}}{2} A_{\frac{3}{2} -\frac{1}{2}} 
-\sin[3(\phi_b-\phi_S)] \, J_3(Qb) \, 
\frac{3}{4} A_{\frac{3}{2} -\frac{3}{2}}
\Big],
\label{eq:dens5}
\end{eqnarray}
where we defined the angle $\phi_b$ in the transverse plane as, 
$\vec b  = b ( \cos \phi_b \hat e_x + \sin \phi_b \hat e_y )$. 
One notices from Eqs.~(\ref{eq:dens4},\ref{eq:dens5}) 
that the transverse charge densities display monopole, dipole, 
quadrupole, and octupole field patterns, which respectively are determined   
by the helicity form factors with zero, one, two, or three 
units of helicity flip between the initial and final $\Delta$ states.

We can now evaluate the electric quadrupole moment corresponding 
to the transverse charge densities $\rho^\Delta_{T \, s_\perp}$.  
Choosing $\vec S_\perp = \hat e_x$, the electric quadrupole 
moment can be defined as~: 
\begin{eqnarray}
Q^\Delta_{s_\perp} \equiv e \int d^2 \vec b \, (b_x^2 - b_y^2) \, 
\rho^\Delta_{T \,  s_\perp}(\vec b). 
\end{eqnarray}
From Eqs.~(\ref{eq:dens4},\ref{eq:dens5}) 
one obtains~:
\be
Q^\Delta_{\frac{3}{2}} = - Q^\Delta_{\frac{1}{2}}
= \frac{1}{2} 
\, \left\{ 2 \left[ G_{M1}(0) - 3 e_\Delta \right] + 
\left[ G_{E2}(0) + 3 e_\Delta \right] \right\} 
\, \left( \frac{e}{M_\Delta^2} \right). 
\label{eq:quadrup}
\ee
We may note that for a spin-3/2 particle without internal structure, 
for which the tree-level predictions are $G_{M1}(0) = 3 e_\Delta$ and $G_{E2}(0) = -3 e_\Delta$, the quadrupole moment of the 
transverse charge densities vanishes.  
It is thus interesting to observe from Eq.~(\ref{eq:quadrup}) 
that $Q^\Delta_{s_\perp}$ is only sensitive to the anomalous parts 
of the spin-3/2 magnetic dipole and electric quadrupole moments, 
and vanishes for a particle without internal structure. The same observation 
was made for the case of a spin-1 particle in Ref.~\cite{Carlson:2008zc}.
Furthermore, the factor 1/2 
multiplying the curly brackets on the 
right hand side of Eq.~(\ref{eq:quadrup}) can be understood by relating the 
quadrupole moment of a 3-dimensional charge distribution 
(which we denote by $Q_{3d}$) 
to the quadrupole moment of a 2-dimensional charge
distribution (denoted by $Q_{2d}$) both defined w.r.t. the spin axis.   
By taking the spin axis along the $x$-axis, the quadrupole moment 
for a 3-dimensional charge distribution $\rho_{3d}$ is defined as~:
\be
Q_{3d} \equiv \int dx dy dz \, ( 3 x^2 - r^2) \, \rho_{3d}(x,y,z), 
= \int dx dy dz \, \left[ (x^2 - y^2) + (x^2 - z^2) \right] \, 
\rho_{3d}(x,y,z) .
\label{eq:q3d}
\ee
For a 3-dimensional charge distribution which is invariant under rotations
around the axis of the spin, the two terms proportional to $(x^2 - y^2)$ 
and $(x^2 - z^2)$ in Eq.~(\ref{eq:q3d}) give equal contributions yielding~:
$
Q_{3d} =2 \int dx dy dz \, ( x^2 - y^2) \, \rho_{3d}(x,y,z).
$
Introducing the 2-dimensional charge density in the $xy$-plane as~:
$
\rho_{2d}(x,y) = \int dz \, \rho_{3d}(x,y,z), 
$
one immediately obtains the relation 
$
Q_{3d} = 2 \, Q_{2d} ,
$
with the quadrupole moment of the 2-dimensional charge density defined as~:
$
Q_{2d} \equiv \int dx dy \, ( x^2 - y^2) \, \rho_{2d}(x,y)$.
Because $Q_{3d}$ is proportional to $G_{E2}(0)$ in our case, we see that 
 $Q_{2d}$, which is half the value of $G_{E2}(0)$, is 
consistent with Eq.~(\ref{eq:quadrup}).

We can also evaluate the electric octupole moment corresponding 
to the transverse charge densities $\rho^\Delta_{T \, s_\perp}$.  
Choosing $\vec S_\perp = \hat e_x$, the electric octupole 
moment can be defined as~: 
\be
O^\Delta_{s_\perp} \equiv e \int d^2 \vec b \, b^3 \, \sin (3 \phi_b) \, 
\rho^\Delta_{T \,  s_\perp}(\vec b)
= e \int d^2 \vec b \, b_y \, (3 b_x^2 - b_y^2) \, 
\rho^\Delta_{T \,  s_\perp}(\vec b).
\ee
From Eqs.~(\ref{eq:dens4},\ref{eq:dens5}) 
one obtains~:
\be
O^\Delta_{\frac{3}{2}} = - \frac{1}{3} O^\Delta_{\frac{1}{2}}
= \frac{3}{2}
\, \Big\{ - G_{M1}(0) - G_{E2}(0) + G_{M3}(0) + e_\Delta  \Big\} 
\, \left( \frac{e}{2 M_\Delta^3} \right). 
\label{eq:octupole}
\ee
We may note that for a spin-3/2 particle without internal structure, 
for which $G_{M1}(0) = 3 e_\Delta$, $G_{E2}(0) = -3 e_\Delta$,  
and $G_{M3}(0) = - e_\Delta$~\cite{Alexandrou:2009hs}
the electric octupole moment of the transverse charge densities 
vanishes.

\section{Lattice Techniques}

The matrix element of the electromagnetic current, $V_\mu$, between two
$\Delta$-states can be decomposed in terms of four independent
covariant vertex function coefficients, $a_1(q^2)$, $a_2(q^2)$, $c_1(q^2)$ and $c_2(q^2)$,
which depend only on the momentum transfer squared $q^2=(p_f-p_i)^2$~\cite{Nozawa:1990gt}:
\begin{eqnarray}
  \langle \Delta(p_f,s_f) |\, V^\mu \, | \Delta(p_i,s_i) \rangle &=& \sqrt{\frac{m_\Delta^2}{E_{\Delta(\vec p_f)}E_{\Delta(\vec p_i)}}} \
                                                \bar u_\sigma(p_f,s_f)\, \Op^{\sigma \mu \tau}\, u_\tau(p_i,s_i) \\  
  \Op^{\sigma\mu\tau} &=& -\delta_{\sigma\tau}                \left[a_1 \gamma^\mu - i \frac{a_2}{2m_\Delta}P^\mu \right] 
                        +\frac{q^\sigma q^\tau}{4m_\Delta^2}\left[c_1 \gamma^\mu - i \frac{c_2}{2m_\Delta}P^\mu \right] \nonumber
   \, .
\end{eqnarray}
$E_\Delta$ and $m_\Delta$ denote the energy and the mass of the particle, $p_{i}\, (p_f)$ and $s_i\, (s_f)$ 
are the initial (final) four-momentum and spin-projection, while $P=p_f+p_i$. 
Every vector-component of the Rarita-Schwinger spinor $u_\sigma(p,s)$ satisfies the free
Dirac equation. Furthermore, two auxiliary conditions are obeyed:
$
   \gamma_\sigma u^\sigma(p,s) = 0$ and 
$   p_\sigma u^\sigma(p,s) = 0 \, .
$
The vertex function coefficients are linked to the phenomenologically more
interesting multipole form factors $G_{E0}$, $G_{E2}$, $G_{M1}$ and $G_{M3}$ by
a linear relation~\cite{Nozawa:1990gt}. The dominant form factors are the electric
charge, $G_{E0}$, and the magnetic dipole, $G_{M1}$, form factors.

\noindent
The  interpolating field 
$
   {\mathbf \chi}^{\Delta^+}_{\sigma\alpha}(x) = \frac{1}{\sqrt{3}} \epsilon^{abc}\Bigl[2\left({\mathbf u}^{a\top}(x) C\gamma_\sigma {\mathbf d}^b(x)\right) {\mathbf u}_\alpha^c(x)  
                                     + \left({\mathbf u}^{a\top}(x) C\gamma_\sigma {\mathbf u}^b(x)\right) {\mathbf d}_\alpha^c(x)\Bigr]\, ,$ has the $\Delta^+$ quantum numbers
where $C$ is the charge conjugation matrix.
To facilitate ground-state dominance  a covariant Gaussian smearing~\cite{Alexandrou:1992ti} on the
quark-fields entering ${\mathbf \chi}^{\Delta^+}_{\sigma\alpha}(x) $ is
used: $
        {\mathbf q }_\beta(t,\vec x) = \sum_{ \vec y} [\eins + \alpha H(\vec x,\vec y; U)]^n \ q_\beta(t,\vec y)$, 
	$H(\vec x, \vec y; U)  = \sum_{\mu=1}^3 \left(U_\mu(\vec x,t)\delta_{\vec x, \vec y - \hat \mu} + U^\dagger_\mu(\vec x-\hat \mu, t) \delta_{\vec x,\vec y+\hat\mu} \right)
$.
Here $q$ is the local quark field (i.e. either $u$ or $d$), ${\bf q}$ is the smeared quark field and $U_\mu$ 
is the $SU(3)$-gauge field. 
For the lattice spacing and pion masses considered in this work, the values
$\alpha=4.0$ and $n=50$ ensure ground state dominance with the shortest
time evolution that could  be achieved.

\subsection{Correlation functions}
We specialize to a kinematical setup where the final $\Delta$-state is at rest ($\vec p_f =\vec 0$) and 
measure the two-point and three-point functions
\begin{eqnarray}
   G_{\sigma \tau}(\Gamma^\nu,\vec p, t_f-t_i) &=&\sum_{\vec x_f} e^{-i\vec x_f \cdot \vec p}\, 
   \Gamma^\nu_{\alpha'\alpha}\, \langle {\mathbf \chi}_{\sigma\alpha}(t_f,\vec x_f) \bar{\mathbf \chi}_{\tau\alpha'}(t_i, \vec 0) \rangle \label{twopoint} \\
   G_{\sigma\ \tau}^{\ \mu}(\Gamma^\nu,\vec q, t) &=& \sum_{\vec x,\, \vec x_f} e^{i\vec x \cdot \vec q}\,
   \Gamma^\nu_{\alpha'\alpha}\, \langle {\mathbf \chi}_{\sigma\alpha}(t_f,\vec x_f) V^\mu(t,\vec x) \bar{\mathbf \chi}_{\tau\alpha'}(t_i, \vec 0)\rangle \, , \label{threepoint}
\end{eqnarray}          
where $V_\mu$ is  the electromagnetic current, which for Wilson fermion is
taken to be the symmetrized, lattice conserved current. We work 
with a representation of the Clifford-algebra in which $\gamma_4$ is diagonal. In this
representation 
$
   \Gamma^k = \frac{1}{2}\left(\begin{array}{l l}\sigma^{(k)} & 0\\ 0 & 0\end{array}\right) \quad {\rm and}\quad \Gamma^4 = \frac{1}{2}\left(\begin{array}{l l}\eins & 0\\ 0 & 0\end{array}\right)\, ,
$
with $k=1,\ldots , 3$ and $\sigma^{(k)}$ being the Pauli matrices.
The ratio
\begin{equation}
        R_{\sigma\ \tau}^{\ \mu}(\Gamma,\vec q,t) = \frac{G_{\sigma\ \tau}^{\ \mu}(\Gamma,\vec q,t)}{G_{k k}(\Gamma^4,\vec 0, t_f)}\ 
				         \sqrt{\frac{G_{kk}(\Gamma^4,\vec p_i, t_f-t)G_{kk}(\Gamma^4,\vec 0  ,t)G_{kk}(\Gamma^4,\vec 0,t_f)}
					            {G_{kk}(\Gamma^4,\vec 0, t_f-t)G_{kk}(\Gamma^4,\vec p_i,t)G_{kk}(\Gamma^4,\vec p_i,t_f)}}\, ,
\end{equation}
with implicit summations over the indices $k$ with $k=1,\ldots, 3$,
becomes time independent for large Euclidean time separations $t_f-t$ and $t-t_i$:
\small
\be
   R_{\sigma\ \tau}^{\ \mu}(\Gamma,\vec q,t) \to \Pi_{\sigma\ \tau}^{\ \mu}(\Gamma,\vec q) =
   \sqrt{\frac{3}{2}}\left[\frac{2 E_{\Delta(\vec q)}}{m_\Delta} 
                          +\frac{2 E^2_{\Delta(\vec q)}}{m^2_\Delta} 
                          +\frac{  E^3_{\Delta(\vec q)}}{m^3_\Delta} 
                          +\frac{  E^4_{\Delta(\vec q)}}{m^4_\Delta} \right]^{-\frac{1}{2}} 
    {\rm tr}\left[\Gamma\, \Lambda_{\sigma\sigma'}(p_f) \Op^{{\sigma'}\mu{\tau'}} \Lambda_{\tau'\tau}(p_i) \right] \, .
\ee
\normalsize
The traces act in spinor-space and the Euclidean Schwinger-Rarita spin sum is given by
\be
   \Lambda_{\sigma\tau}(p)  = - \frac{-i\slashed{p}+m_\Delta}{2m_\Delta}\left[
                               \delta_{\sigma\tau}-\frac{\gamma_\sigma\gamma_{\tau}}{3}
                               +\frac{2p_\sigma p_{\tau}}{3m_\Delta^2} 
			       - i \frac{p_\sigma\gamma_{\tau}-p_{\tau}\gamma_\sigma}{3m_\Delta} \right] \, .
\ee

Since we are evaluating the correlator of Eq.~(\ref{threepoint}) using  
sequential inversions through the sink~\cite{Dolgov:2002zm}, 
a separate set of inversions is necessary for every choice of vector and Dirac-indices.
The total of $256$ combinations is beyond our computational resources, and hence we concentrate on a few 
carefully chosen combinations given by
\be
   \Pi_\mu^{(1)}(\vec q) = \sum \limits_{j,k,l=1}^3 \epsilon_{jkl}\Pi_{j\ k}^{\ \mu}(\Gamma^4, \vec q), \quad
   \Pi_\mu^{(2)}(\vec q) = \sum \limits_{k=1}^3 \Pi_{k\ k}^{\ \mu}(\Gamma^4, \vec q),  \quad
   \Pi_\mu^{(3)}(\vec q) = \sum \limits_{j,k,l=1}^3 \epsilon_{jkl}\Pi_{j\ k}^{\ \mu}(\Gamma^j, \vec q) \label{comb}\, .
\ee
From these all the multipole form factors can be optimally  extracted. For instance
the first relation in Eq.~(\ref{comb}) 
is proportional to $G_{M1}$, while the third isolates $G_{E2}$ for $\mu=4$.

\subsection{Data analysis}
For a given value of $q^2$ the combinations given in Eqs.~(\ref{comb})  are evaluated
for all different directions of $\vec q$ resulting in the same $q^2$, as well as for 
all four directions $\mu$ of the current. This leads to an over-constrained linear
system of equations, which is then solved in the least-squares sense yielding 
estimates of $G_{E0}$, $G_{E2}$, $G_{M1}$ and $G_{M3}$. The estimates are 
embedded into a jackknife binning procedure, 
thus providing  statistical 
errors for the form factors that take all correlation and 
autocorrelation effects
into account. 

The details of the simulations are 
summarized in Table~\ref{bigtab}.
In each case, the separation between the final and
initial time is $t_f-t_i \gtrsim 1\,{\rm fm}$ and Gaussian smearing is
applied to both source and sink to suppress contamination from higher states having the  quantum numbers of the $\Delta(1232)$. 
For the mixed-action calculation, the 
domain-wall valence quark mass was chosen  to reproduce
the lightest pion mass obtained using $N_F=2+1$
improved staggered quarks~\cite{Edwards:2005ym, Hagler:2007xi}.
\begin{table}[h]
\begin{center}
\caption{\label{bigtab}Lattice parameters and results.  $N_{\rm conf}$ denotes the
number of lattice configurations, $\sqrt{\langle r^2\rangle}$ gives the charge radius,  $\mu_{\Delta^+}$ is the $\Delta^+$ magnetic moment in nuclear
magnetons and $Q^\Delta_{\frac{3}{2}}$ is the $\Delta^+$ quadrupole moment. }
\begin{tabular}{ c c c c c c}
 $N_{\rm conf}$ & $m_\pi$ [GeV] & $m_\Delta$ [GeV] & $\sqrt{\langle r^2\rangle}$ [fm] & $\mu_{\Delta^+}$ [$\mu_N$]& $Q^\Delta_{\frac{3}{2}}$ \\
\hline
\multicolumn{6}{c}{Quenched Wilson, $32^3\times 64$, $a=0.092$~fm}\\
\hline
 200       & 0.563(4)      & 1.470(15)        &    0.6147(66)       &   1.720(42)   &  0.96(12)  \\
 200       & 0.490(4)      & 1.425(16)        &    0.6329(76)       &   1.763(51)   &  0.91(15)        \\
 200       & 0.411(4)      & 1.382(19)        &    0.6516(87)       &   1.811(69)   &  0.83(21)           \\
\hline
\multicolumn{6}{c}{$N_F=2$ Wilson, $24^3\times 40 (32$ for lightest pion), $a=0.077$~fm 
}\\
\hline
 185       & 0.691(8)      & 1.687(15)        &  0.5279(61)       &   1.462(45) & 0.80(21)            \\
157       & 0.509(8)       & 1.559(19)        &  0.594(10)        &   1.642(81) &  0.41(45)       \\
 200       & 0.384(8)      & 1.395(18)        &  0.611(17)        &   1.58(11)  & 0.46(35)      \\
\hline
\multicolumn{6}{c}{$N_F=2+1$, Mixed action, $28^3\times 64$, $a=0.124$~fm \cite{Aubin:2004wf}}\\
\hline
 300       & 0.353(2)      & 1.533(27)        &  0.641(22)          &    1.91(16)   & 0.74(68)          \\
\end{tabular}
\end{center}
\end{table}

\section{Results}
\begin{figure}[H]
\begin{minipage}{7.cm}
\vspace*{-0.3cm}
\includegraphics[width =7.5cm,height=7.5cm]{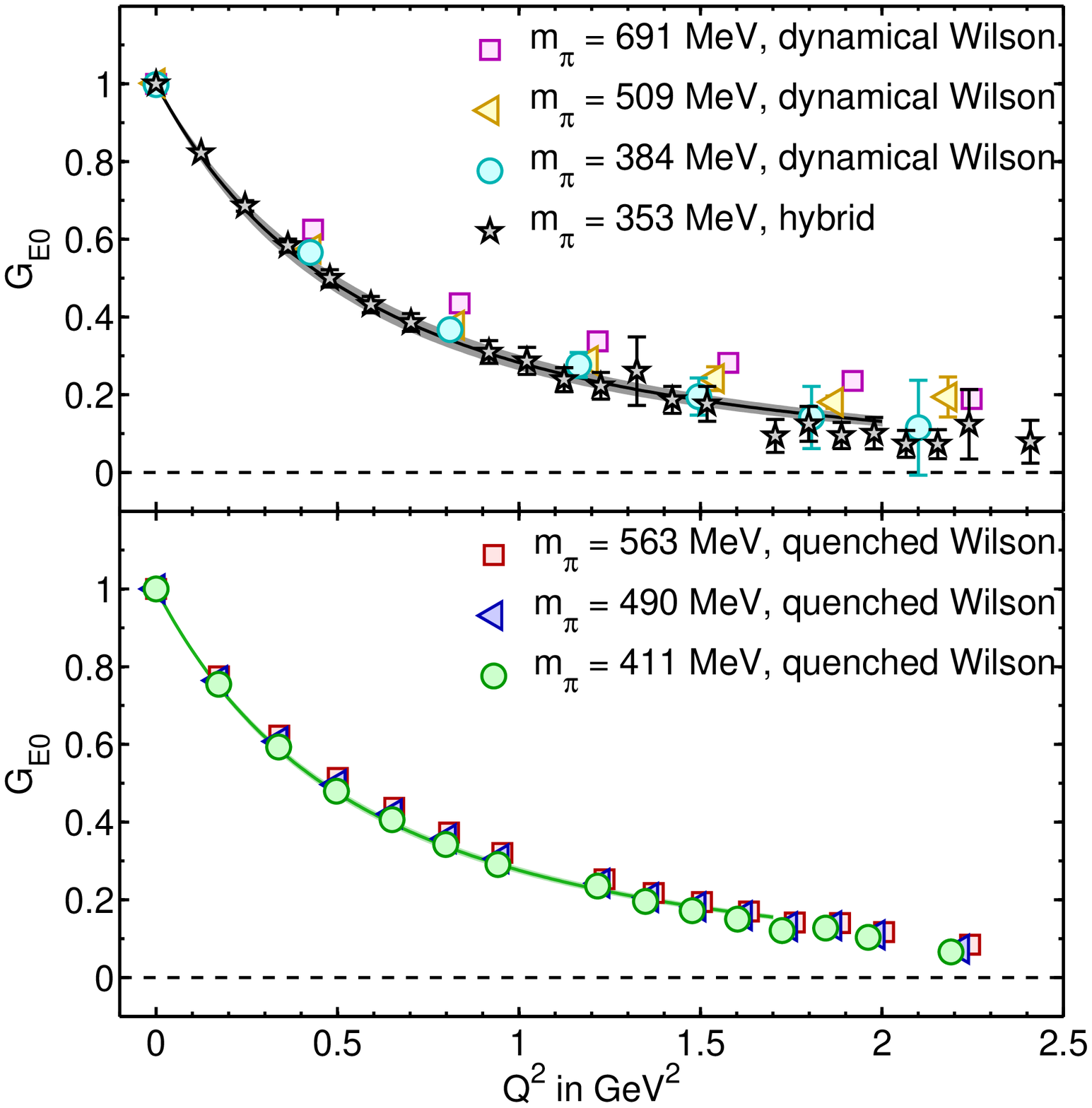}
\caption{
Comparison of the three  different QCD lattice calculations for the 
$\Delta^+(1232)$  form factor
 $G_{E0}$. The upper curve shows a dipole fit to results using
the mixed action. Results are also
shown for two dynamical Wilson fermions.
 The lower curve is a dipole fit to results in the quenched 
approximation~\cite{Alexandrou:2007we}. The error bands are calculated using a jackknife analysis of the fit parameters
}
\label{fig:ge0}
\end{minipage}
\hfill
\begin{minipage}{7.cm}
\vspace*{-2.cm}
\includegraphics[width =7.5cm,height=7.5cm]{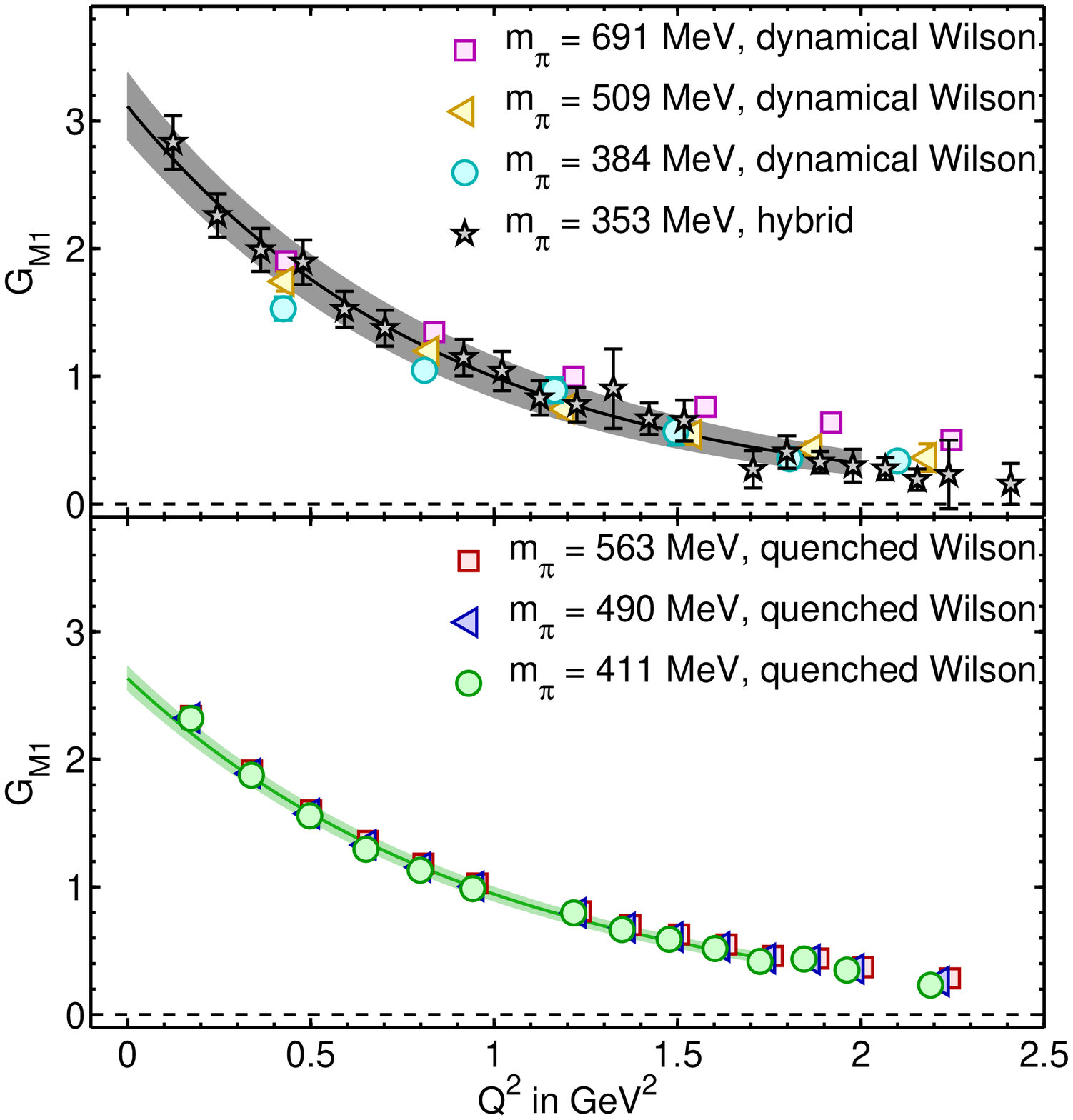}
\caption{
Comparison of the three different QCD lattice calculations for the 
$\Delta^+(1232)$ magnetic dipole form factors
 $G_{M1}$. The lines show the fits to an exponential
 form. The rest of the notation is the same as that in Fig.~1.
}
\label{fig:gm1}
\end{minipage}
\end{figure}

\begin{figure}[H]
\begin{minipage}{7cm}
\includegraphics[width =7.5cm,height=7.5cm]{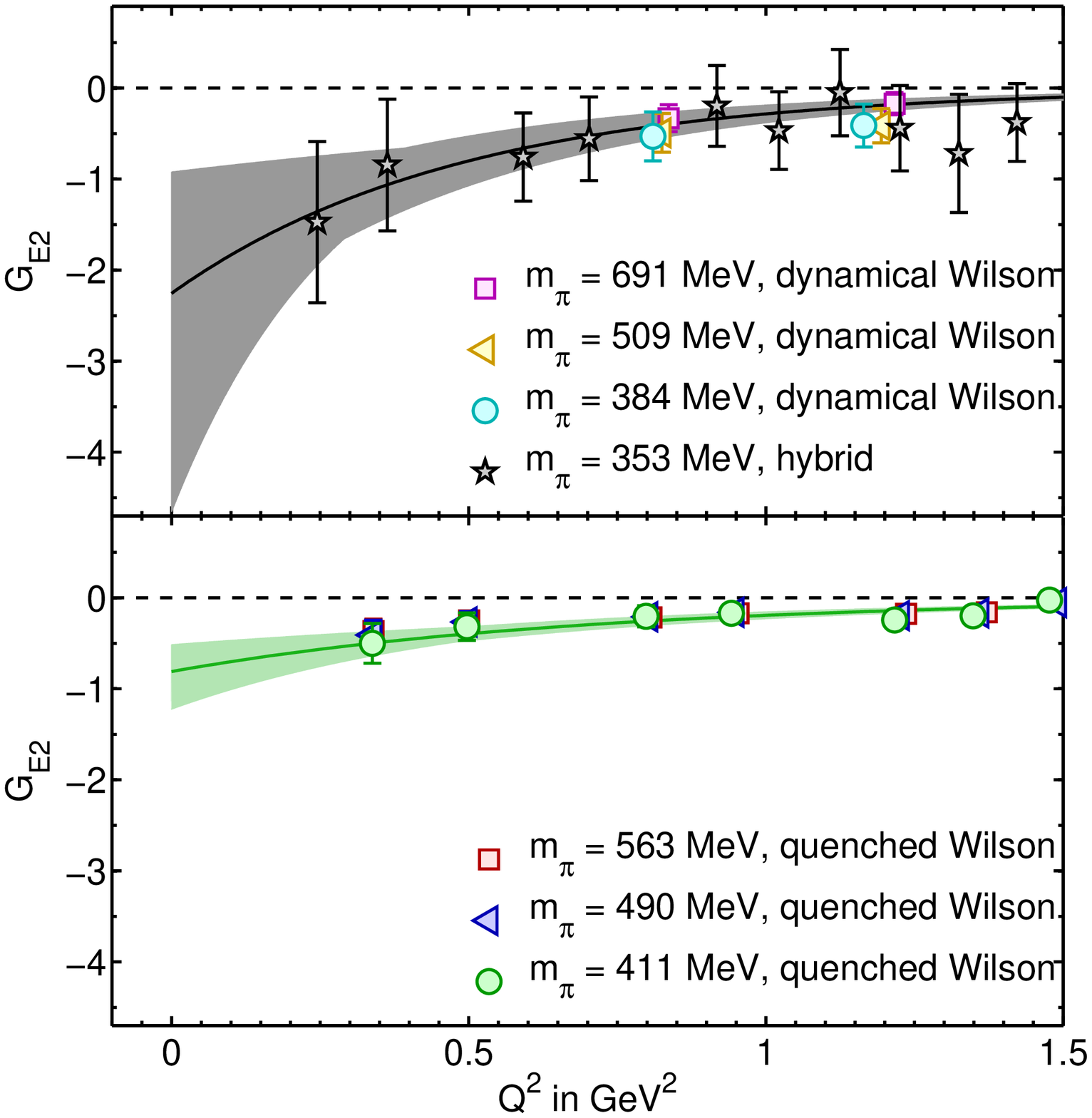}
\caption{
Comparison of the three different QCD lattice calculations for the 
$\Delta^+(1232)$ electric quadrupole form factors
 $G_{E2}$.  The notation is the same as that in Fig.~1.
}
\label{fig:ge2}
\end{minipage}
\hfill
\begin{minipage}{7cm}
\includegraphics[width =7.5cm,height=5cm]{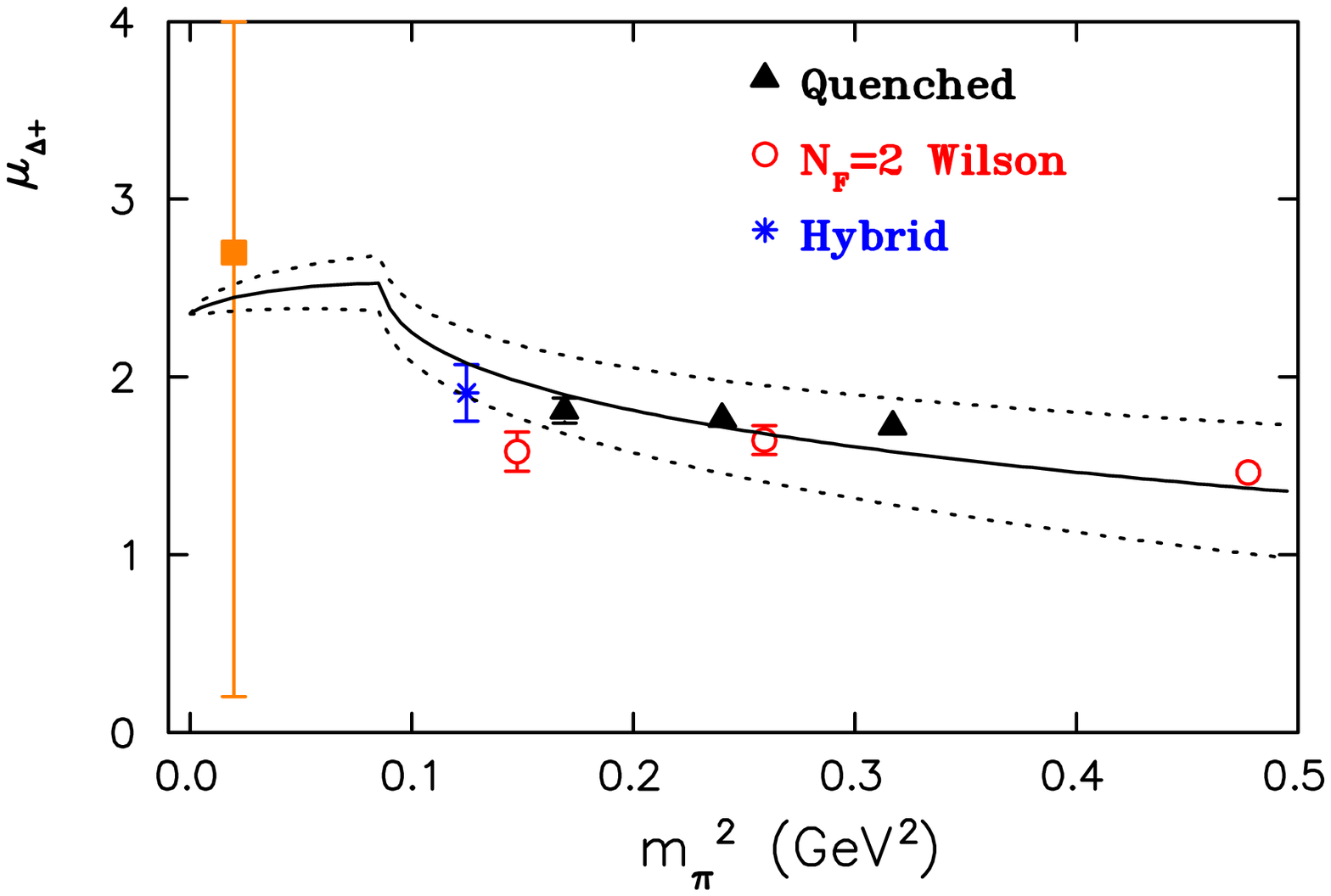}
\caption{The magnetic dipole moment in nuclear magnetons. The value at the physical pion mass (filled square) is shown with  statistical and systematic
errors~\cite{Kotulla:2002cg}.  The solid and dashed curves  show the
results of ChEFT  with the  theoretical error estimate~\cite{Pascalutsa:2004je} .}
\label{fig:mu}
\end{minipage}
\end{figure}

In Figs~\ref{fig:ge0}, \ref{fig:gm1} and \ref{fig:ge2} we show 
lattice results for $N_F=0$ and $N_F=2$ Wilson fermions
and in the mixed action. For the pion masses considered in this work 
there is agreement among results using different actions,
with statistical errors being smallest in the quenched theory.

\begin{figure}[H]
\hspace*{-0.5cm}
\includegraphics[width = 0.3 \linewidth]{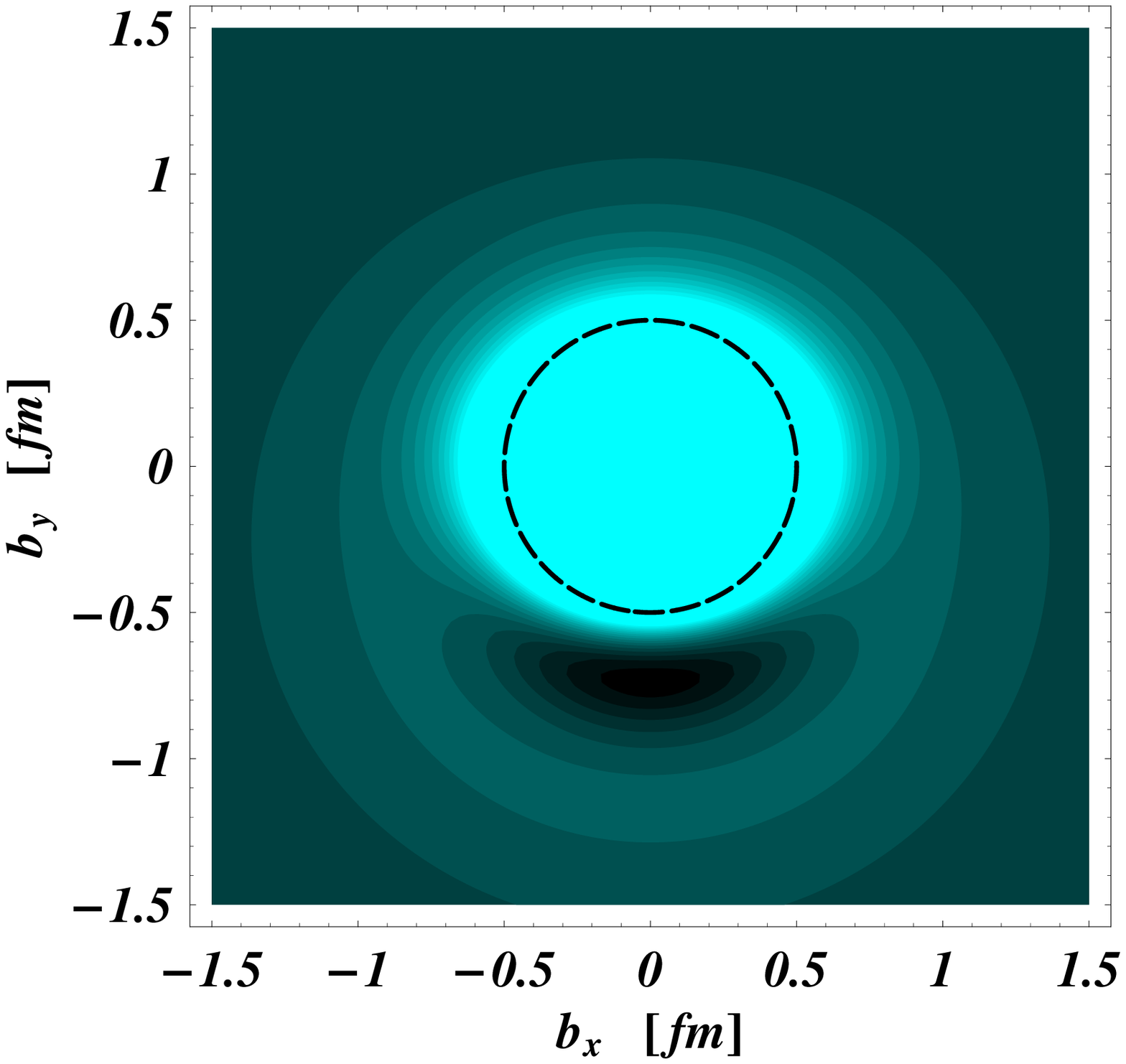}
\includegraphics[width = 0.3 \linewidth]{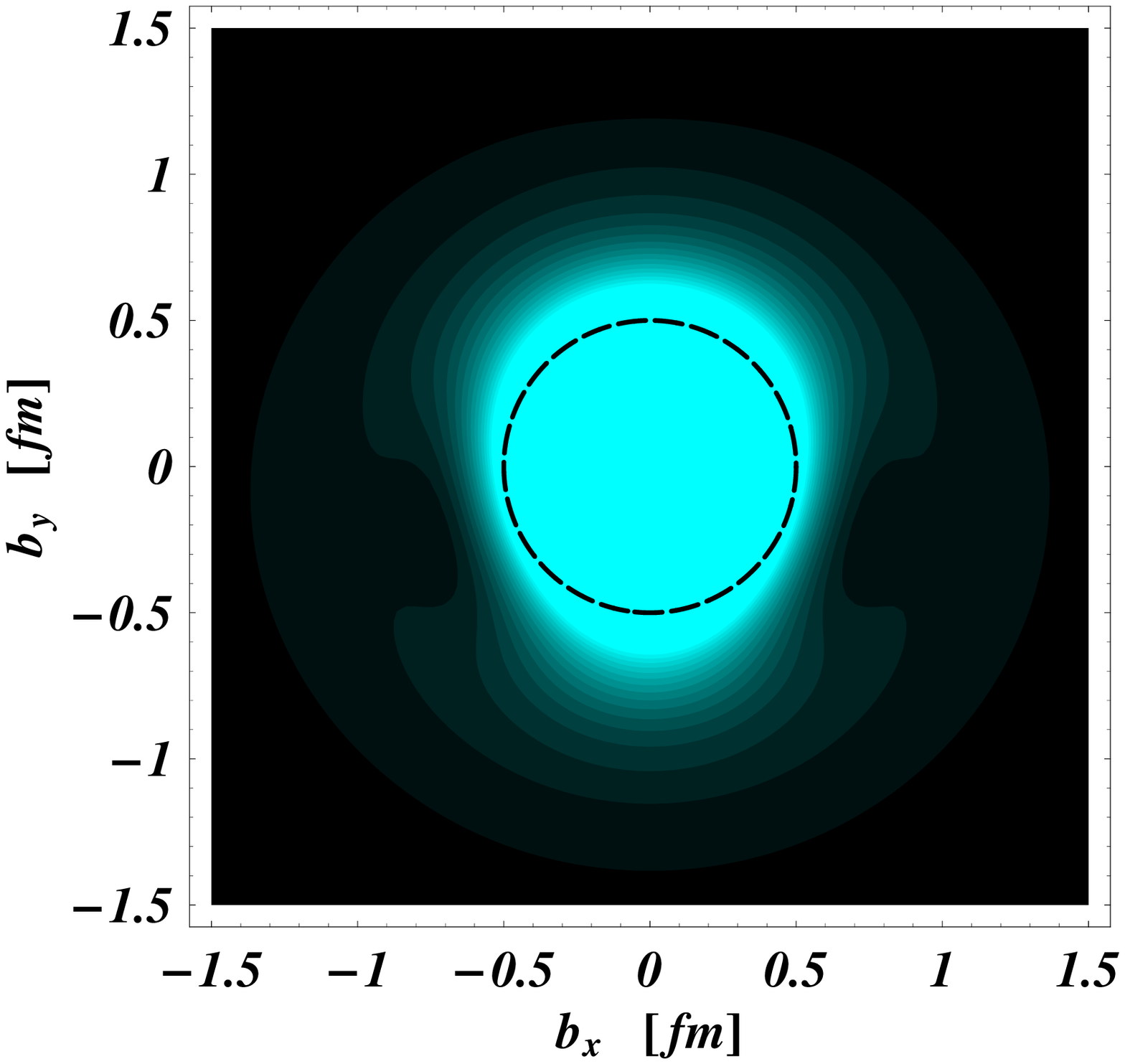}
\includegraphics[width = 6cm, height=5cm]{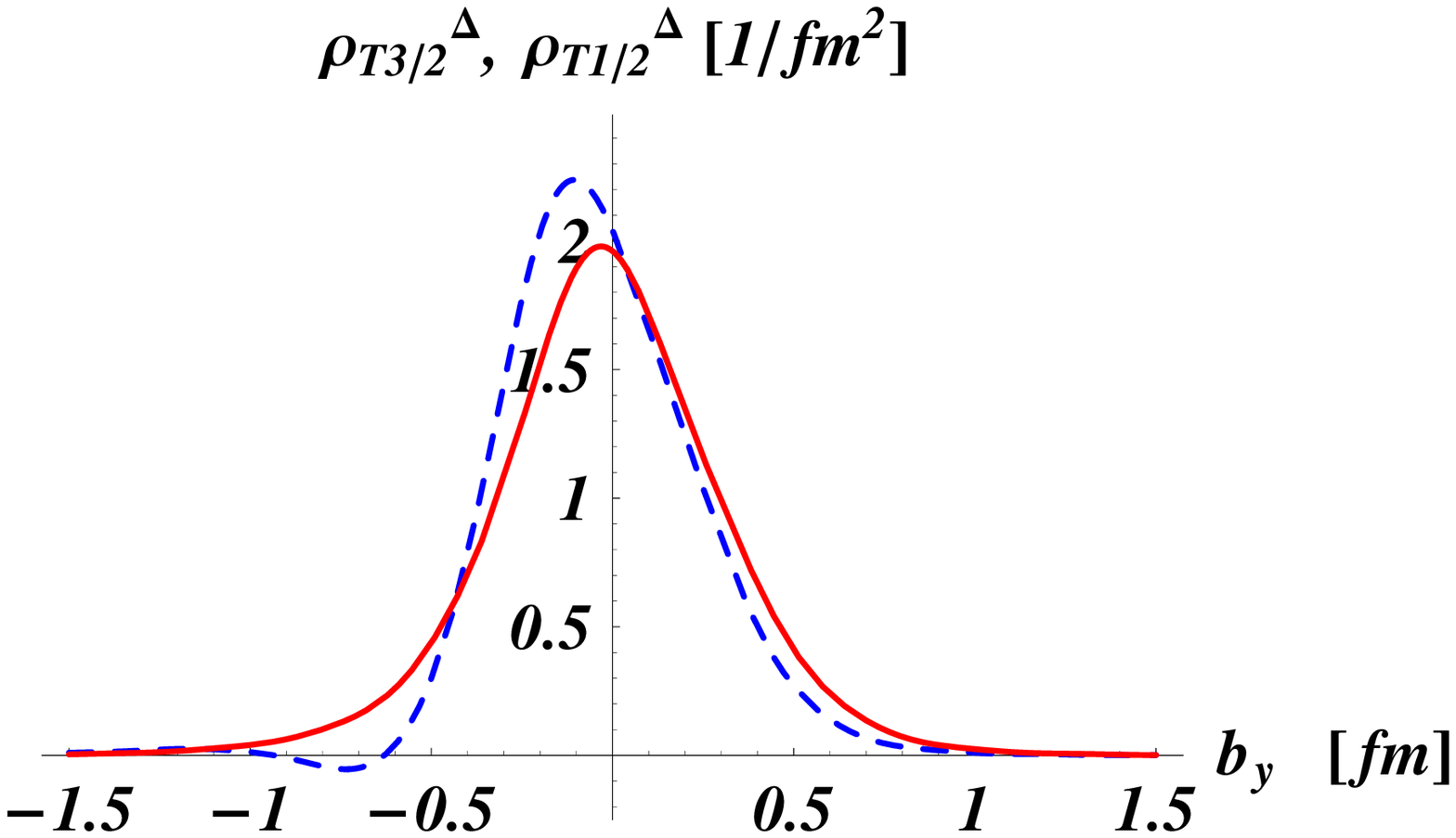}
%
\caption{Quark transverse charge densities in a 
{\it $\Delta^+(1232)$} which 
is polarized along the positive $x$-axis. 
Left panel : $\rho^\Delta_{T \, \frac{3}{2}}$. 
Middle panel : $\rho^\Delta_{T \, \frac{1}{2}}$. 
The light (dark) regions correspond with
largest (smallest) values of the density. 
To better see the deformation, a circle of radius 0.5~fm is drawn for 
comparison. Right panel compares the density along the $y$-axis 
for  $\rho^\Delta_{T \, \frac{3}{2}}$ (dashed curve) and 
$\rho^\Delta_{T \, \frac{1}{2}}$ (solid curve). 
For the $\Delta$ e.m. FFs, 
the quenched lattice QCD results are used.}
\label{fig:deltatrans}
\end{figure}

The magnetic moment as function of $m_\pi^2$ is shown in Fig.~\ref{fig:mu},
together with a comparison to a chiral effective field theory (ChEFT) result~\cite{Pascalutsa:2004je}.  
The  ChEFT result has one free parameter (a low-energy constant) that
has been fitted to lattice data, shown by the central line. 
We also estimate the uncertainty of the ChEFT expansion (expansion in 
pion mass) by the error band in Fig.~\ref{fig:mu}. The uncertainty of the 
ChEFT calculation vanishes in the chiral limit because in this limit one  
simply has the value of the low-energy constant (LEC), which lies 
within the broad experimental error 
band $\mu_{\Delta^{+}}= 2.7^{+1.0}_{-1.3}(stat.)\pm 
1.5 (syst.) \pm 3.0 (theory)\mu_N$~\cite{Kotulla:2002cg}. 
In this work, we do not consider the uncertainty 
in the fit value of the LEC due to the lattice errors, 
as the calculations are still performed for pion masses 
where the $\Delta$ is stable (on the right side of the kink).
A calculation for pion mass values where the $\Delta$ becomes unstable will be 
a challenge for future calculations. 
 The $\Delta$ moments using an approach similar to ours
 are calculated only in the  quenched approximation~\cite{Leinweber:1992hy,Zanotti:2004jy,Boinepalli:2006ng}.
Our magnetic moment results agree  with recent 
background field calculations using dynamical improved Wilson
fermions~\cite{Aubin:2008hz}.
The spatial length $L_s$ of our lattices
satisfies $L_s m_\pi> 4 $ in all cases 
except at the lightest pion mass with $N_F=2$ Wilson fermions, for which 
$L_s m_\pi =3.6$.  For that point,  the magnetic moment falls slightly below the error band, consistent with the fact shown in  Ref.~\cite{Aubin:2008hz}
 that finite
  volume effects  decrease the magnetic moment.

\begin{figure}[H]
\hspace*{-0.5cm}
\includegraphics[width = 0.32 \linewidth]{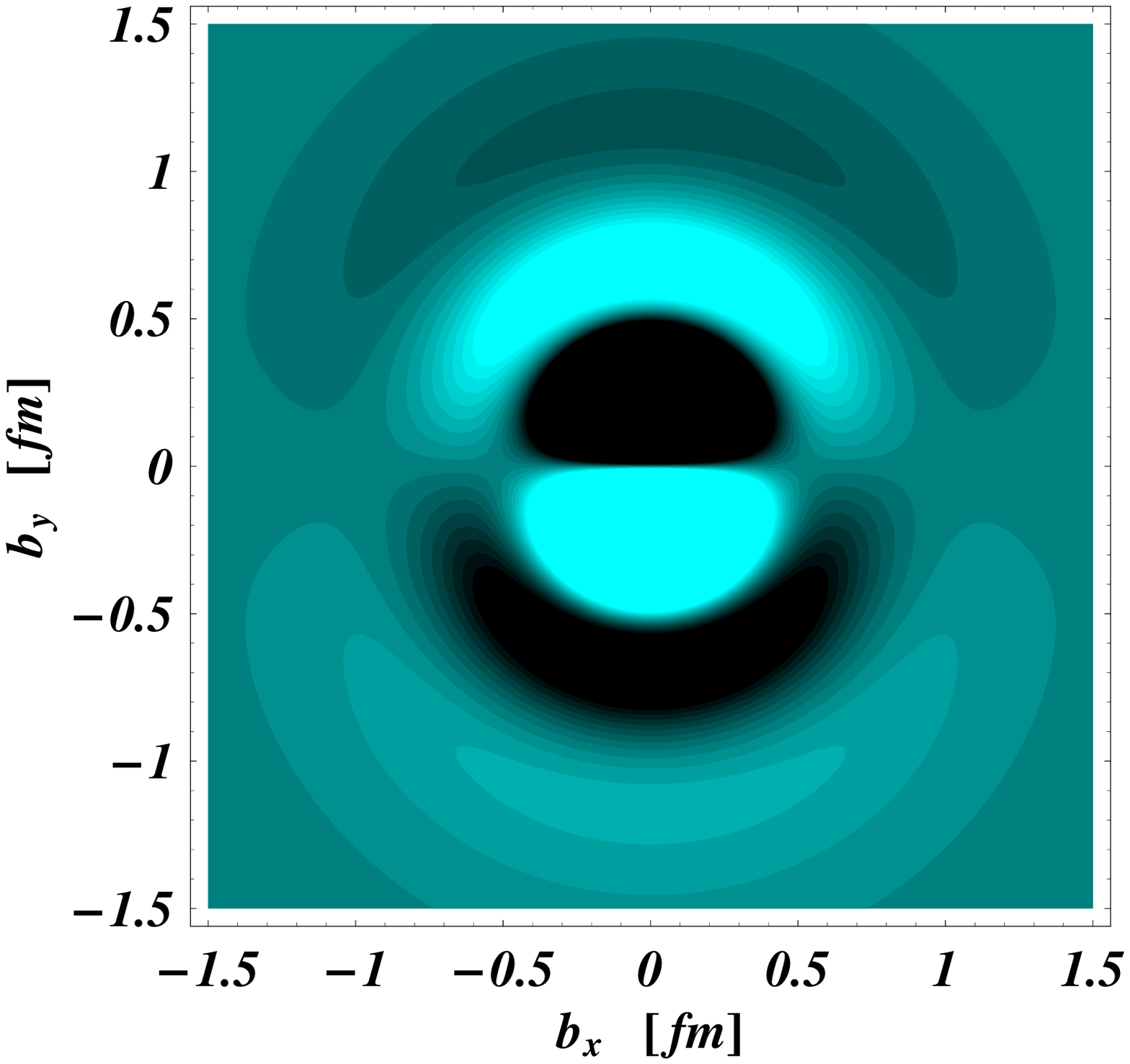}
\includegraphics[width = 0.32 \linewidth]{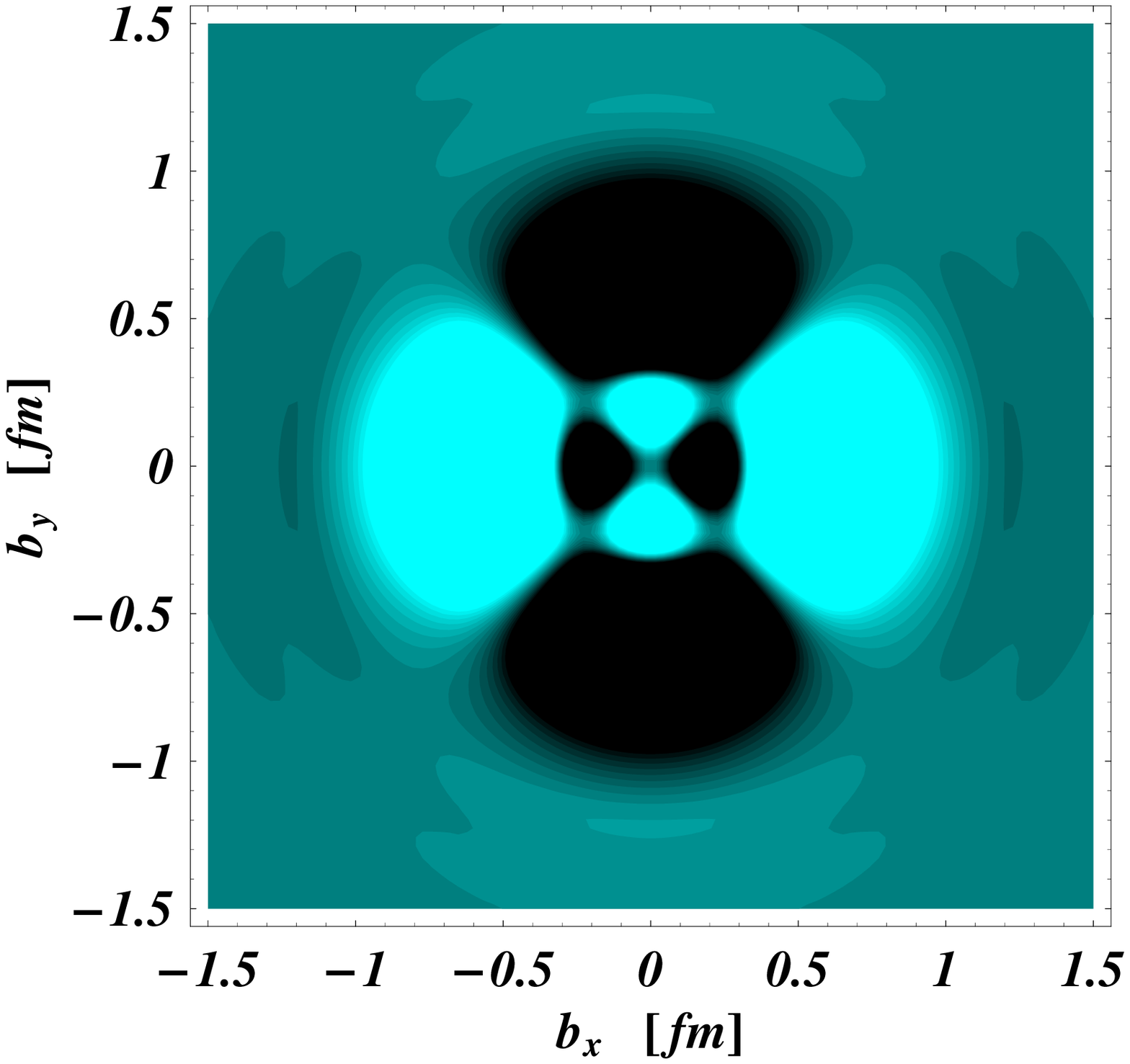}
\includegraphics[width = 0.32 \linewidth]{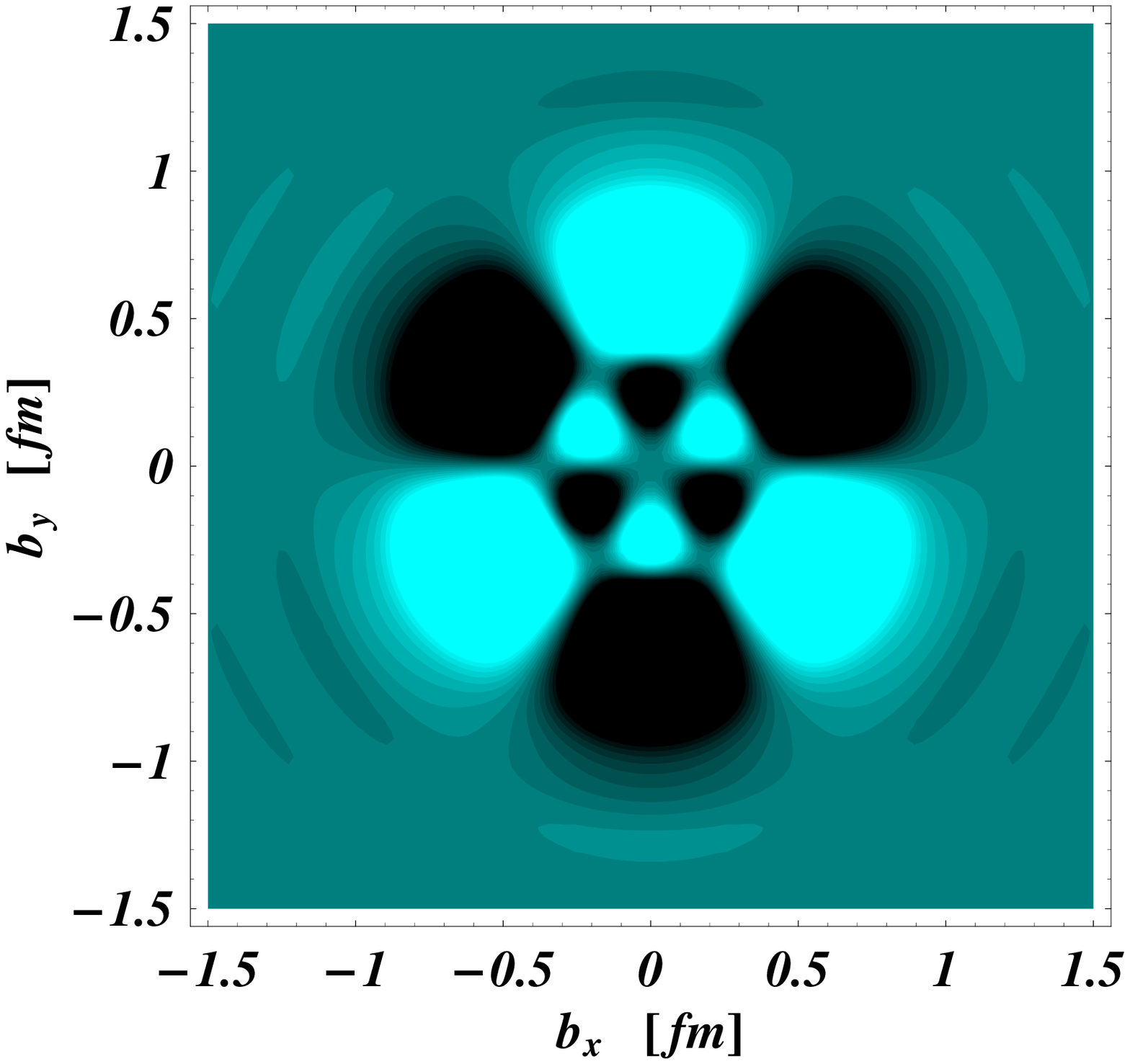}
\hspace{0.25cm}
\caption{Different field patterns in the 
quark transverse charge density $\rho^\Delta_{T \, \frac{3}{2}}$ 
in a {\it $\Delta^+(1232)$} which 
is polarized along the positive $x$-axis. 
Left panel : dipole field pattern; 
Middle panel : quadrupole field pattern; 
right panel : octupole field pattern.  
For the $\Delta$ e.m. FFs,  
the quenched lattice QCD results are used.} 
\label{fig:deltatrans2}
\end{figure}

In Fig.~\ref{fig:deltatrans}, the transverse densities of 
Eqs.~(\ref{eq:dens4},\ref{eq:dens5}) are compared 
for a $\Delta^+$ which has a transverse spin. It is seen that the 
quark charge density in a 
$\Delta^+$ in a state of transverse spin projection 
$s_\perp = +3/2$ is elongated 
along the axis of the spin (prolate deformation)  
whereas in a state of transverse spin 
projection $s_\perp = +1/2$ it is elongated along the axis perpendicular to
the spin.
In Fig.~\ref{fig:deltatrans2} the dipole, quadrupole and octupole
 field patterns 
for the $\Delta$ transverse densities  are clearly seen.

\section{Conclusions}
\label{sec8}

We have presented a study of  the electromagnetic 
properties of the $\Delta(1232)$-resonance using lattice QCD.
The lattice results for the $\Delta$  electromagnetic
form factors have been shown here down to approximately 350 MeV for  
three cases:
quenched QCD, two flavors of dynamical Wilson quarks, and three  
flavors of
quarks described by a mixed action combining domain wall valence quarks
and  staggered sea quarks.

We have also worked out the formalism which allows to interpret
these results in terms of the quark charge densities.
More specifically,
we have established the relation between the light-front helicity  
amplitudes
and form factors for the case of electromagnetic interaction of a  
spin-3/2 particle,
which allowed us to obtain the quark transverse
charge densities of the $\Delta$ using lattice results for the form  
factors.

The light-front formalism
allows for a consistent relation between the internal structure
of the particle and its shape.
Our lattice results show that the quark charge density in the
$\Delta^+$ is elongated along the axis of the spin. 

\noindent
{\bf Acknowledgments:} This work is supported in part by the Cyprus Research Promotion Foundation  under contracts EPYAN/0506/08
and $\Delta$IE$\Theta$NH/$\Sigma$TOXO$\Sigma$/0308/07.

\end{document}